\documentclass[aps,
               reprint,
               prb,
               twocolumn,
               superscriptaddress]{revtex4-2}
\usepackage{bm}
\usepackage{graphicx}
\usepackage{graphics}
\usepackage{amsmath}
\usepackage{amsfonts}
\usepackage{amssymb}
\usepackage{makeidx}
\usepackage[colorlinks=true,citecolor=blue,linkcolor=blue,linktocpage=true,pagebackref=false]{hyperref}
\hypersetup{colorlinks=true,citecolor=blue,linkcolor=blue,filecolor=blue,urlcolor=blue}
\usepackage{color,soul} 
\usepackage{float}
\usepackage{subfigure}

\usepackage{xcolor}

\usepackage{siunitx}

\graphicspath{{figures/}} 
\begin{document}

\newcommand{\be}   {\begin{equation}}
\newcommand{\ee}   {\end{equation}}
\newcommand{\ba}   {\begin{eqnarray}}
\newcommand{\ea}   {\end{eqnarray}}
\newcommand{\ve}   {\varepsilon}
\newcommand{\Dis}  {\mbox{\scriptsize dis}}

\newcommand{\state} {\mbox{\scriptsize state}}
\newcommand{\band} {\mbox{\scriptsize band}}

\title{Quasicrystalline 30$^{\circ}$ twisted bilayer graphene: \\ 
fractal patterns and electronic localization properties}

\author{Kevin J. U. Vidarte}
\affiliation{Instituto de F\'{\i}sica, Universidade Federal
  do Rio de Janeiro, 21941-972 Rio de Janeiro - RJ, Brazil}
\author{Caio Lewenkopf}
\affiliation{Instituto de F\'{\i}sica, Universidade Federal Fluminense, 24210-346 Niter\'oi - RJ, Brazil}

\date{13 September 2024}
\begin{abstract}
The recently synthesized 30$^\circ$ twisted bilayer graphene (30$^\circ$-TBG) systems are unique quasicrystal systems possessing dodecagonal symmetry with graphene's relativistic properties. 
We employ a real-space numerical atomistic framework that respects both the dodecagonal rotational symmetry and the massless Dirac nature of the electrons to describe the local density of states of the system. 
The approach we employ is very efficiency for systems with very large unit cells and does not rely on periodic boundary conditions. 
These features allow us to address a broad class of multilayer two-dimensional crystal with incommensurate configurations, particularly TBGs.
Our results reveal that the 30$^\circ$-TBG electronic spectrum consist of extended states together with a set of localized wave functions.
The localized states exhibit fractal patterns consistent with the quasicrystal tiling. 
\end{abstract}
\maketitle

\section{Introduction}
\label{sec:introduction}

The electronic properties of bilayer systems when twisted at specific angles, have been a subject of intense research interest.
In particular, twisted bilayer graphene (TBG) near to the so-called magic angles show a plethora of phenomena, such as superconductivity \cite{Cao2018, Yankowitz2019, Lu2019}, magnetism \cite{Sharpe2019, Lin2022, Jaoui2022, Vidarte2024}, and strong electronic correlation effects \cite{Cao2018a, Lu2019}. 
Some of these interesting phenomena are also present in twisted multilayer graphene 
\cite{Brzhezinskaya2021,Kononenko2022}.
Recently, the discovery of dodecagonal quasicrystals \citep{Yao2018,Ahn2018} in $30^{\circ}$-twisted bilayer graphene ($30^{\circ}$-TBG) has further attracted considerable attention to TBGs, as it offers a unique platform to study the effects arising from long-range order without translational symmetry.  
Unlike conventional quasicrystals, where all atoms are intrinsically located within a quasiperiodic order \citep{Shechtman1984, Levine1984}, a $30^{\circ}$-TBG is viewed as an extrinsic quasicrystal because its quasiperiodicity arises from the interlayer coupling between two graphene monolayers \cite{Moon2019}.

Dodecagonal $30^{\circ}$-TBG quasicrystals have been successfully synthesized on various substrates, including SiC \cite{Ahn2018, Bocquet2020, Fukaya2021}, Pt \cite{Yao2018}, Cu \cite{Yan2019, Pezzini2020, Deng2020, Liu2021}, and a Cu-Ni alloy \cite{Takesaki2021}, using methods such as chemical vapor deposition and carbon segregation. 
The $30^{\circ}$-TBGs are experimentally identified by low-energy electron diffraction (LEED) \citep{Ahn2018,Yao2018,Bocquet2020,Takesaki2021}, high-energy positron diffraction (HEPD)\citep{Fukaya2021}, transmission electron microscopy (TEM) \citep{Ahn2018,Deng2020}, scanning tunneling microscopy (STM) \citep{Yan2019}, and Raman spectroscopy \citep{Yao2018,Pezzini2020,Liu2021}. 
Furthermore, magnetotransport measurements have confirmed the low-energy Dirac fermionic nature of these systems \citep{Yan2019,Pezzini2020}.

Various techniques have been employed to characterize dodecagonal quasicrystalline properties of $30^{\circ}$-TBGs.
Angle-resolved photoemission spectroscopy (ARPES) measurements have revealed multiple Dirac cones with 12-fold rotational symmetry in $30^{\circ}$-TBGs \citep{Yao2018,Ahn2018}, 
a feature that is highly sensitive to deviations from the $30^{\circ}$ twist angle \cite{Ahn2018}.
The emergence of mirrored Dirac cones within the Brillouin zone of each graphene layer and a bandgap at the zone boundary, attributed to generalized Umklapp scattering between the two layers, characterizes their coupling \cite{Yao2018, Ahn2018}.
STS measurements \citep{Yan2019} reveal a suppression of density of state (DOS) consistent with the bandgap observed in previous ARPES studies, suggesting the same origin \citep{Yan2019}.
Additionally, time- and angle-resolved photoemission spectroscopy measurements show an unbalanced electron distribution in the replica Dirac cone bands \cite{Suzuki2019}.

The dodecagonal $30^{\circ}$-TBG quasicrystal is a direct realization of the 
Stampfli construction process \cite{Stampfli1986}, which involves equilateral triangles, squares, and rhombuses as fundamental units arranged in a dodecagonal pattern. 
False-colored TEM images of $30^{\circ}$-TBGs \citep{Ahn2018,Deng2020}
clearly depict the Stampfli tiles in various orientations. 

Atomic-resolved STM images of $30^{\circ}$-TBGs \cite{Yan2019} have offered insights into the unique electron density distribution in real space. 
A striking feature is the flower-like contrast observed on the ``coronene" patches of graphene, which exhibits higher symmetry than adjacent regions. 
These ``flowers" align with the typical patterns of the dodecagonal quasicrystal tiling elements.

The lack of translation invariance in $30^{\circ}$-TBGs poses significant challenges for numerical calculations of its electronic properties.
To address this, several strategies have been employed.
One approach involves a series of periodic approximations with minimal lattice mismatch between the two layers to effectively reproduce the electronic and optical properties of $30^{\circ}$-TBG within a finite unit cell involving more than $10^7$ atoms \citep{Yu2019}. 
Another proposed effective method is a $\textbf{k}$-space tight-binding model 
\citep{Moon2019, Yu2020_2}.
The model reveals wave functions with characteristic fractal inflations of the dodecagonal quasicrystal tiling.
Moreover, an inverse participation ratio (IPR) analysis shows that the system
12-fold symmetric resonant states are localized \cite{Moon2019}, a hallmark of quasicrystals \cite{Niu1986, Kohmoto1987}. 
These states are energetically far from the charge neutrality point, but it has been theoretically shown that by applying pressure or an electric field they can be pushed towards the Fermi energy \cite{Yu2020_1}.

Recent theoretical predictions for $30^{\circ}$-TBG systems are rich and diverse. 
Numerical simulations suggest that the perfect superlubricity characterized by a scale-invariant sliding force related to the contact area, can be achieved for geometric sequences of dodecagonal quasicrystal tiling elements \citep{Koren2016}.
The system's quasiperiodicity and weak interlayer coupling can originate quantum oscillations with spiral Fermi surfaces \citep{Spurrier2019}. 
We note that the nature of interlayer hybridization selection rules governing interlayer coupling has been thoroughly investigated, via symmetry and group representation theory \citep{Yu2022}. 
Unlike the only allowed equivalent hybridization in $D_{6h}$ untwisted graphene bilayer, $D_{6}$ twisted graphene bilayer permits equivalent and mixed hybridizations.
Furthermore, $D_{6d}$ dodecagonal quasicrystal in $30^{\circ}$-TBG allows for both equivalent and nonequivalent hybridizations \citep{Yu2022}.
Additionally, the presence of an exchange field and a Rashba spin-orbit coupling have been proposed to induce a nontrivial topological phase in $30^{\circ}$-TBG systems \citep{Li2020}.

In this study, we reexamine the electronic properties of $30^{\circ}$-TBGs, with a focus on the spatial distribution of the localized states, using 
the Haydock-Heine-Kelly (HHK) recursion technique \citep{Haydock1972, Haydock1975, Haydock1980, Vidarte2022}.
This method relies on the nearsightedness of electronic matter (see, for instance, Ref.~\cite{Prodan2005}) and, as a real-space approach, unlike the standard techniques \citep{Martin2004}, eliminates the need for periodic boundary conditions, making it ideal for analyzing strongly disordered, amorphous, and quasicrystalline electronic systems.
We demonstrate that the HHK method outperforms the previously proposed approaches for $30^{\circ}$-TBG systems.
By developing a geometric model to interpret the LDOS results, we are able to uncover the spatial distribution of the LDOS suppression observed in Ref.~\cite{Yao2018} and gain insights into the fractal nature of the localized electronic states. 

\section{Theory and methods}
\label{sec:method_and_methods}

\subsection{Lattice structure}
\label{sec:atomic_structure}
 
We define the dodecagonal quasicrystal structure for a $30^{\circ}$-TBG system by initiating from AA-stacked bilayer graphene \cite{CastroNeto2009} and subsequently rotating the top layer by $30^{\circ}$ around the shared hexagonal center of both layers.
The primitive lattice vectors of the bottom layer are given by $\textbf{a}^{b}_{1}=\sqrt{3} a_0 \hat{\bf e}_{x}$ and $\textbf{a}^{b}_{2}=\sqrt{3}a_0/2 \left( \hat{\bf e}_{x} + \sqrt{3}\hat{\bf e}_{y} \right)$, while the 
the top layer ones are defined by rotating the bottom layer vectors, namely, $\textbf{a}^{t}_{i}=R(\pi/6)\textbf{a}^{b}_{i}$.
Here, the interlayer spacing between graphene layers is $d_{0} = 3.35$~\AA~ and the carbon-carbon bond length is $a_{0}=1.42$~\AA~ \citep{CastroNeto2009}.

The obtained $30^{\circ}$-TBG system exhibits a dodecagonal quasicrystal structure with $D_{6d}$ point group symmetry, consistent with experimental observations \citep{Yu2022}. 
The system's symmetry is primarily defined by six reflection planes $\sigma_{d}$ and six two-fold rotation axes $C'_{2}$. 
Carbon atoms equidistant from the twist axis are equivalent and related through the application of these symmetry operators.

The $30^{\circ}$-TBG lattice can be spatially mapped onto a quasicrystal lattice model constructed by Stampfli tiles \citep{Stampfli1986} discussed below.
Here, we discuss the spatial distribution of lattice sites and propose a geometrical model based on two distinct Stampfli quasicrystal lattice construction methods. 

\begin{figure}[h]
\centering
\includegraphics[width=0.9\columnwidth]{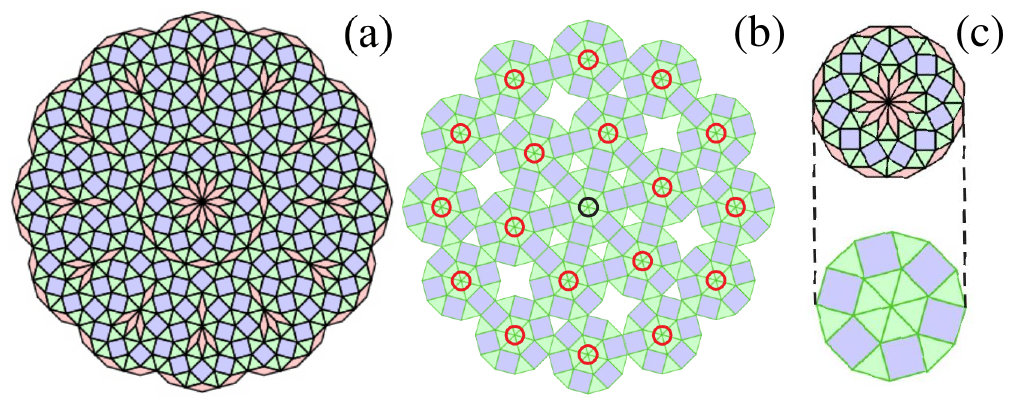}
\caption{Stampfli quasicrystal construction methods. (a) First construction method used a grid with 12-fold rotational symmetry consisting of equilateral triangles, squares and rhombuses.
(b) Second construction method as a compound tessellation consisting of equilateral triangles and squares. 
The red and black circles indicate the dodecagonal centers.
(c) Two dodecagons of identical size for both Stampfli construction methods.
}
\label{fig:Fig_01}
\end{figure}

Figure \ref{fig:Fig_01}(a) illustrates the first Stampfli construction method obtained from the grid method \citep{Stampfli1986,Ahn2018}.
The quasicrystal lattice is composed of equilateral triangles, squares and rhombuses (with $30^{\circ}$ acute angle), arranged with different orientations to maintain a 12-fold rotational symmetry pattern, while lacking translational symmetry.
The inflation rule of Stampfli tiles involved rotating the lattice by $15^{\circ}$ around its perfect rotational symmetry center and scaling side lengths by a factor $\sqrt{2+\sqrt{3}}$ \citep{Stampfli1986}.
This process generates a new lattice whose vertices coincide with the ones of the original lattice.

Figure \ref{fig:Fig_01}(b) presents the second construction method as a compound tessellation \citep{Koren2016,Stampfli1986}.
The dodecagonal compound tessellation is decomposed into equilateral triangles and squares, as depicted in Fig.~\ref{fig:Fig_01}(c).
The center of each dodecagon corresponds to a vertex of a larger, similar dodecagonal element, with side lengths scaled by a factor $2+\sqrt{3}$ \citep{Stampfli1986}.

In the next section, we use the properties of the Stampfli construction methods to discuss electronic emergent localization features of $30^{\circ}$-TBG systems.
More specifically, we employ the third inflation of Stampfli tiles, seed multiplied by a factor $\left( 2+\sqrt{3} \right)^{3/2}$, as used in the first construction method. 
This ensures that the internal structures of each tile kind are similar \citep{Ahn2018}.

\begin{figure}[h]
\centering
\includegraphics[width=0.85\columnwidth]{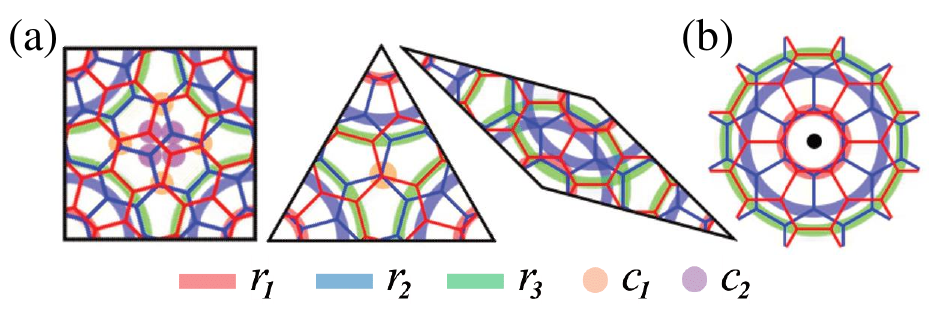}
\caption{Geometric model: 
The five different highlighted-color, marked as $r_{1}$, $r_{2}$, $r_{3}$, $c_{1}$, $c_{2}$, indicate an enhanced concentration of carbon atom sites at five different radial distances, using as references 
(a) the Stampfli tile vertices and (b) the dodecagonal centers obtained by the first and second construction methods, respectively. 
Red and blue lines correspond to the top and bottom graphene layers, respectively.
}
\label{fig:Fig_02}
\end{figure}

Let us introduce a geometric model that identifies the main locations of enhanced concentration of lattice sites. 
Figure \ref{fig:Fig_02} highlights five distinct radial distances denoted as $r_{1}$, $r_{2}$, $r_{3}$, $c_{1}$ and $c_{2}$, where the Stampfli tile vertices serve as central points of emergent geometric site concentration.
Central points formed by a single kind of Stampli tiles, coinciding with the main twist axis perpendicular to the plane, exhibit a perfect 12-fold symmetry.
In contrast, central points formed by the vertices of different Stampfli tiles, corresponding to carbon atom sites at the same radial distances, are characterized by imperfect rings.
Most imperfect rings exhibit an imbalance in emergent state measures, as detailed in Sec. \ref{sec:results}.
The imbalance primarily originates from the distinct symmetry groups of the Stampfli tiles: rhombuses, triangles and squares possess twofold, threefold and fourfold symmetry, respectively, see Fig.~\ref{fig:Fig_02}(a).
Near-perfect rings are readily identified by the dodecagonal centers used in the second construction method,  red circles in Fig.~\ref{fig:Fig_01}(b), where the center point regions exhibit pronounced 12-fold symmetry and balanced emergent state measures.

In summary, we use two complementary lattice construction methods to study electronic localization properties of $30^{\circ}$-TBGs.
The first construction method effectively identifies all emergent localization measures, particularly those with exhibiting imbalances.
The second construction method makes it easier to identify emergent localization measures with pronounced 12-fold symmetry.

\subsection{Model Hamiltonian}
\label{sec:model_hamiltonian}

The low-energy electronic structure of an incommensurate 30$^\circ$-TBG system is effectively described by the single-orbital tight-binding Hamiltonian \citep{CastroNeto2009,Reich2002}
\be
\label{eq:H_tb}
H = \sum_{i>j} \left(  t_{ij} c^\dagger_i c^{}_j + {\rm H. c} \right) ,
\ee
where $i$ and $j$ represent the lattice points and $t_{ij}$ is the transfer integral between the Wannier $p_z$ electronic orbitals centered at the carbon atoms at the sites $i$ and $j$. 
The key parameter in this model is the transfer integral  $t_{ij} \equiv t({\mathbf R}_{ij})$, with ${\mathbf R}_{ij} ={\bf R}_i - {\bf R}_j$, where $\textbf{R}_{i}$ and $\textbf{R}_{j}$ are lattice vectors associated with the sites $i$ and $j$.
The hopping term $t_{ij}$ is cast as \citep{Laissardiere2010,Moon2013,Uryu2004}
\be
\label{eq:Hopping_energy}
t({\mathbf R}) = V_{pp\pi} (R)  \left[ 1-\left( \frac{ {\mathbf R} \cdot  \mathbf{e}_z } {R} \right)^2 \right] +  V_{pp\sigma} (R) \left( \frac{ {\mathbf R} \cdot  \mathbf{e}_z } {R} \right)^2\!,
\ee
with
\begin{align}
     V_{pp\pi} (R) =  V^0_{pp\pi} \exp \left( -\frac{R - a_0}{\delta} \right) , \\
 V_{pp\sigma} (R) =  V^0_{pp\sigma} \exp \left( -\frac{R - d_0}{\delta} \right),
\end{align}
where $V^{0}_{pp\pi}=-2.7$~eV is the transfer integral between the nearest-neighbor atoms that belong to the same graphene layer, $V^{0}_{pp\sigma}=0.48$~eV is the interlayer transfer integral between atoms located at different graphene sheets, and $\delta=0.45$~\AA~ is a characteristic decay length of the transfer integral \citep{Uryu2004,Laissardiere2010}. 
For $R > 5$~\AA~ the transfer integral is exponentially small and can be safely neglected.
The intra- and interlayer transfer integrals have been obtained from a fit of the dispersions of graphene monolayer and graphene AB-stacked bilayer calculated by density functional theory methods \citep{Laissardiere2010}.

Variations in the interlayer coupling strength, $V^{0}_{pp\sigma}$, result in a monotonically shift of the energy peaks corresponding to the largest spectral weight of the atomic sites with the strongest interlayer interaction. 
These states are located at energies far from the Fermi level. 
On the other hand, variations in the intralayer hopping parameter, $V^{0}_{pp\pi}$, primarily affect the Fermi velocity. 
The combined influence of both the intralayer and interlayer hopping parameters determines the energy
position of the density of states suppression arising from the intersection of two Dirac cones, as
discussed in Ref.~\cite{Yao2018}.

The influence of an external magnetic field $\textbf{B}$ is accounted for by the Peierls substitution \citep{Peierls1933, Cresti2021}.
This approach modifies the hopping integrals between lattice sites to account for the magnetic field effect on electron motion, namely,
\be 
\label{eq:Peierls}
t_{ij} \longrightarrow t_{ij} \, \exp \left[ -i \dfrac{e}{\hslash} \int_{\textbf{R}_{i}}^
{\textbf{R}_{j}} d\textbf{r}\cdot \textbf{A}(\textbf{r}) \right] ,
\ee
where 
$\hbar$ is the Planck constant and $-e$ is the electron charge. 
We choose the vector potential as $\textbf{A}(\textbf{r})= \left( 0,Bx,0\right)$, that gives a constant magnetic field $\textbf{B}=\nabla\times \textbf{A} = B\hat{\bf e}_{z}$ perpendicular to TBG.

\subsection{Numerical method}
\label{sec:numerical_method}

We compute the local spectral function of $30^{\circ}$-TBG quasicristals, described by Eqs.~\eqref{eq:H_tb}-\eqref{eq:Peierls}, using the Haydock-Heine-Kelly (HHK) recursion technique \citep{Haydock1972, Haydock1975, Haydock1980, Vidarte2022}.
The HHK method is an $O(\mathcal{N})$ approach which consists of a recursive procedure \citep{Haydock1980} that transforms an arbitrary sparse Hamiltonian matrix in a tridiagonal form.
Next, one evaluates the diagonal Green's function, $G_{ii}^{r}(\epsilon) \equiv G^{r}(\textbf{R}_{i},\textbf{R}_{i},\epsilon)$, by a continued fraction expansion, that is very amenable for numerical calculation.
A detailed presentation of the method and its implementation can be found,
for instance, in Refs.~\citep{Haydock1980,Vidarte2022}.

The local density of states (LDOS) at any site $j$ can be written as
\begin{align}
\label{eq:LDOS}
\rho_{j}(\epsilon)&= -\frac{1}{\pi} {\rm Im}\;G^{r}_{jj}(\epsilon) \\ \nonumber
&\equiv -\frac{1}{\pi} \lim_{\eta \rightarrow 0^{+}} \left[ {\rm Im}\;G_{jj}(\epsilon+i\eta) \right] .
\end{align} 
In practice, a finite $\eta$ serves as a convenient regularization parameter \citep{Vidarte2022,Haydock1975}.
The latter is related to physical processes when addressing realistic systems, since it accounts for the imaginary part of the Green's function self-energy due, for instance, to disorder effects that are ubiquitous in graphene systems \citep{Mucciolo2010}.

As noted in the introduction, the the HHK method is an order $N$ real-space approach that does not require periodic boundary conditions periodic boundary conditions or finite system sizes.
Here, we increase the number of considered lattice sites until the LDOS converges,
typically with approximately $N \sim 10^7$ atomic sites.
Although, in principle, the calculation must be repeated for each atomic site, the geometric model introduced above, combined with the self-similarity of the electronic spectral functions for different lattice inflations (see below), indicates that only a finite number of sites need to be considered.
This feature is essential for the computational efficiency of the method.  

\section{Results}
\label{sec:results}

Using the HHK method we compute the LDOS given by Eq.\eqref{eq:LDOS} and confirm that the equivalent sites at identical radial distances share the same local spectral function due to the most relevant symmetry operators, as discussed in Sec.~\ref{sec:atomic_structure}. 
Consequently, our calculations require the computation of only one twenty-forth of the total number of carbon atoms, within a specified circular region of the $30^{\circ}$-TBG dodecagonal quasicrystal.

\begin{figure}[h]
\centering
\includegraphics[width=0.95\columnwidth]{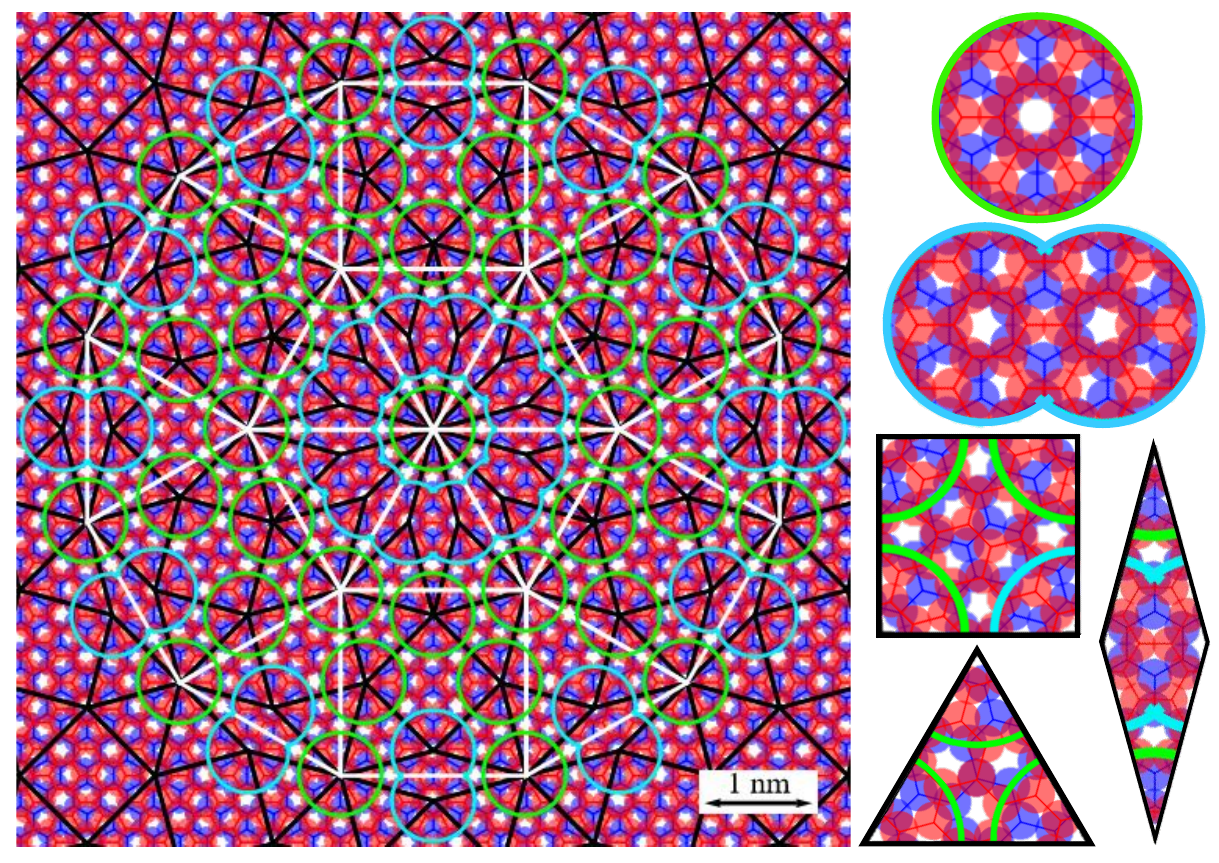}
\caption{Electron density distribution in a real-space image of the $30^{\circ}$-TBG system. 
Black lines represent the third inflation of Stampfli tiles using the first construction method, while the white lines correspond to the second method.
The areas of the red and blue disks are proportional to the local charge density in the top and bottom layers, respectively. 
The purple regions correspond to overlapping blue and red disks.
The green and sky-blue circles indicate the flower-like structure.
The figures on the right side correspond to a zoom-in of the restructure on the left side.
}
\label{fig:Fig_03}
\end{figure}

Figure~\ref{fig:Fig_03} presents a numerical simulation of the electron density distribution at the charge neutrality point, 
$\int_{-\infty}^{\epsilon_{\rm CNP}} d\epsilon\rho_i(\epsilon)$, in a real-space image of the $30^{\circ}$-TBG system.
The real-space image reveals a homogeneous electron density distribution within each graphene layer.
When the two layers are considered together, a non-periodic dodecagonal pattern emerges.
This finding is in line with experimentally obtained atomic resolution STM images \citep{Deng2020,Yan2019}.

The inset in Figure~\ref{fig:Fig_03} reveals an intriguing feature of the 
space-resolved electron density distribution of a $30^{\circ}$-TBG: the flower-like structures. 
These structures indicate local-symmetry regions associated with 12-atom rings.
Collectively, the flower-like structures exhibit a characteristic dodecagonal quasicrystal pattern with strict 12-fold rotational symmetry.
Within each flower-like structure, the local symmetry regions also possess 12-fold rotational symmetry.
In the Stampfli rhombuses, specifically on the smaller diagonal, two flower-like structures are linked by two carbon atoms located in the second ring closest to the center points.

Contrary to what has been described in the experimental literature \citep{Yan2019}, all flower-like structures are easily identified by the first Stampfli construction model, and not by the second one.

\subsection{Local electronic structures}
\label{Local electronic structures}

\begin{figure}[h]
\centering
\includegraphics[width=0.95\columnwidth]{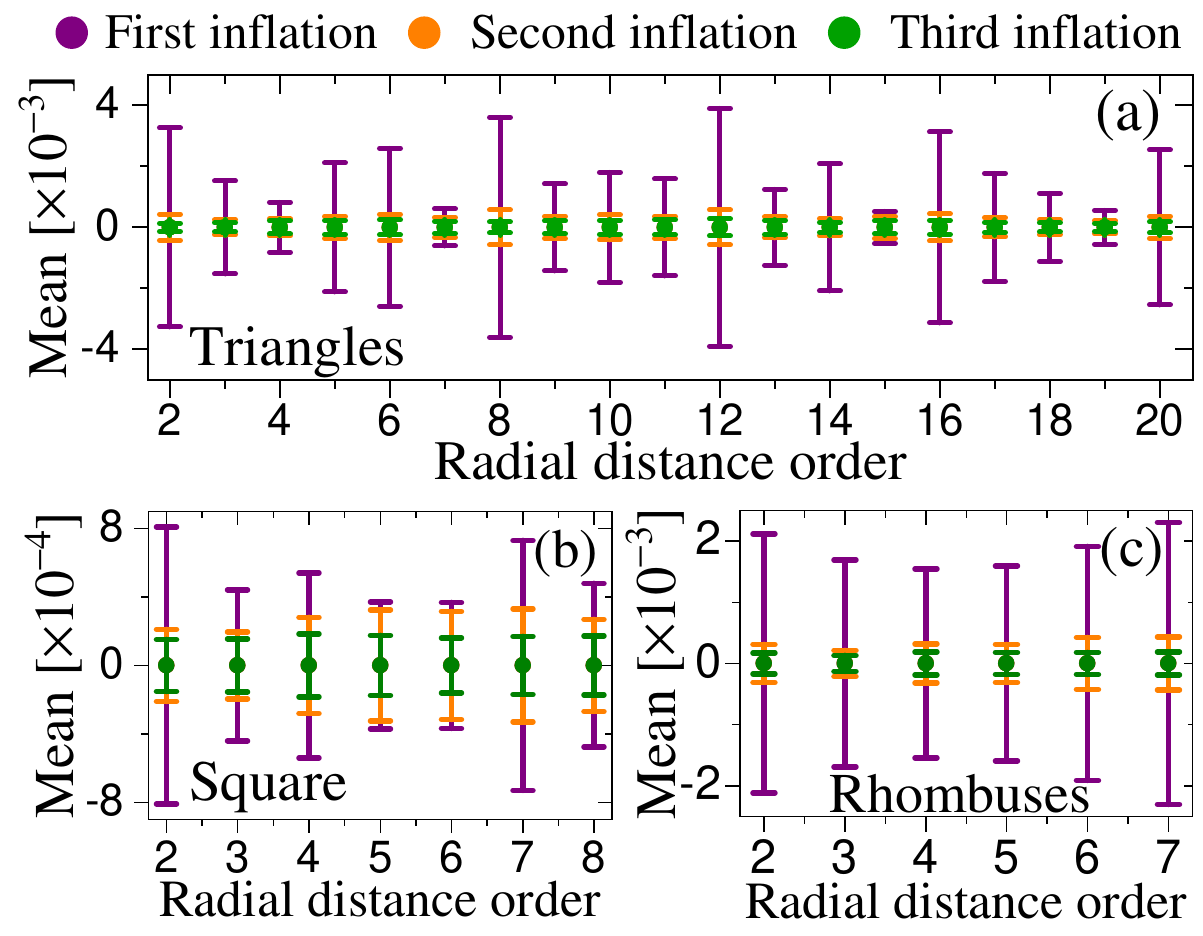}
\caption{
Electronic structure evaluation with different orders of inflation for unit structures such as (a) triangles, (b) squares, and (c) rhombuses.
Means correspond to the sum of discrepancies between two spectral functions over the energy band for different radial distances.
The error bars quantify the standard deviations from the mean value that are mainly due to the singularities of the valence band.
}
\label{fig:Fig_04}
\end{figure}

Figure~\ref{fig:Fig_04} shows the evaluation of ``{\it integrated spectral functions}"  for various Stampfli inflation orders, considering unit structures such as equilateral triangles, squares, and rhombuses.
The integrated spectral function of each Stampfli unit structure is computed as the sum of the set of local spectral functions, $F_{i}(\epsilon)=\sum_{j \in i} \rho_{j}(\epsilon)$, where $j$ labels the carbon atoms sites within the Stampfli tile and $i$ indicates radial distance order.
The averaged deviations between the spectral functions for different inflation orders and the reference spectral function, $F_{1}(\epsilon)$, 
are computed as $\int_{-\infty}^{\infty} \left[F_{i}(\epsilon) - F_{1}(\epsilon)\right]^2 d\epsilon$.
This analysis provides insights into the impact of inflation order on the electronic properties of the Stampfli tiles.

In Fig.~\ref{fig:Fig_04} error bars quantify the standard deviation from the mean value, primarily due to the peaks associated to the valence band singularities that vary along the radial distance.
For a given kind of Stampfli tile, the first inflation shows fluctuations of order $10^{-3}$ and $10^{-4}$ along the radial distance. 
By the third inflation, standard deviations are negligible for all spectral functions, regardless of radial distances.
A small standard deviation indicates greater similarity with the first spectral function.
Thus, the third inflation is the smallest order at which the spectral functions of the Stampfli unit structures exhibit typical patterns characteristic of the dodecagonal quasicrystal.
By accounting for the normalization factor that relates the spectral functions of the different Stampfli unit structures, we observe an excellent agreement within the numerical precision.
Our findings suggest that the electronic structure of the system can be calculated using only a single Stampfli unit structure. 

In summary, the crystalline structure and local spectral functions of $30^{\circ}$-TBGs match the Stampfli tiles, with minor discrepancies observed for the first and second inflations.
These slight differences primarily stem from variations in the internal structures of Stampfli tiles at different radial distances \citep{Ahn2018}.
As the sizes of the Stampfli tiles increase, these differences become increasingly insignificant, revealing the fractal structure of the LDOS.

\subsection{Emergent localization}
\label{Emergent localization}

\begin{figure}[h]
\centering
\includegraphics[width=0.99\columnwidth]{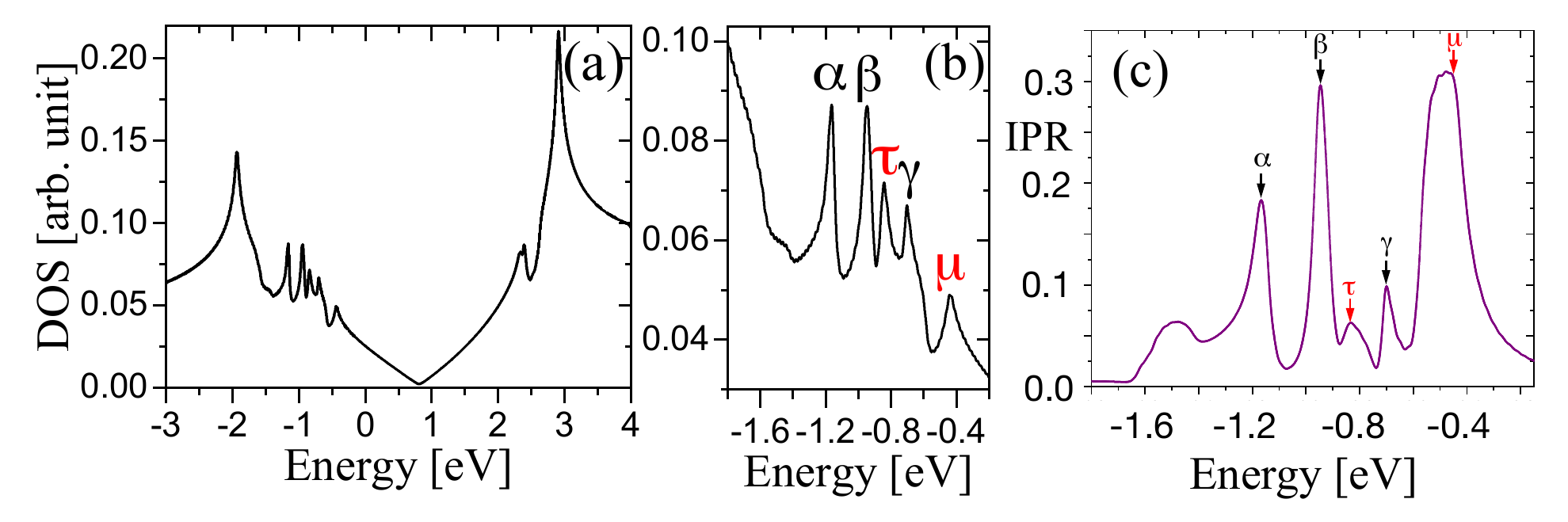}
\caption{
(a) DOS of the dodecagonal quasicrystalline in $30^{\circ}$-TBG. 
(b) Peaks marked as $\alpha$, $\beta$, $\gamma$, $\tau$, $\mu$ correspond to the DOS singularities in the valence band.
(c) IPR for the same energy window.}
\label{fig:Fig_05}
\end{figure}

Let us now examine the localized states corresponding to the singularities in the valence band \cite{Moon2019}.
Figure~\ref{fig:Fig_05} presents the $30^{\circ}$-TBG system density of states, DOS$(\epsilon) = \sum_i \rho_i(\epsilon)$,  and the inverse participation ratio,
IPR$(\epsilon) = \sum_i [\rho_i(\epsilon)]^2$, based on the quasicrystal lattice model constructed by Stampfli tiles.

\begin{figure*}
\centering
\includegraphics[width=0.95\textwidth]{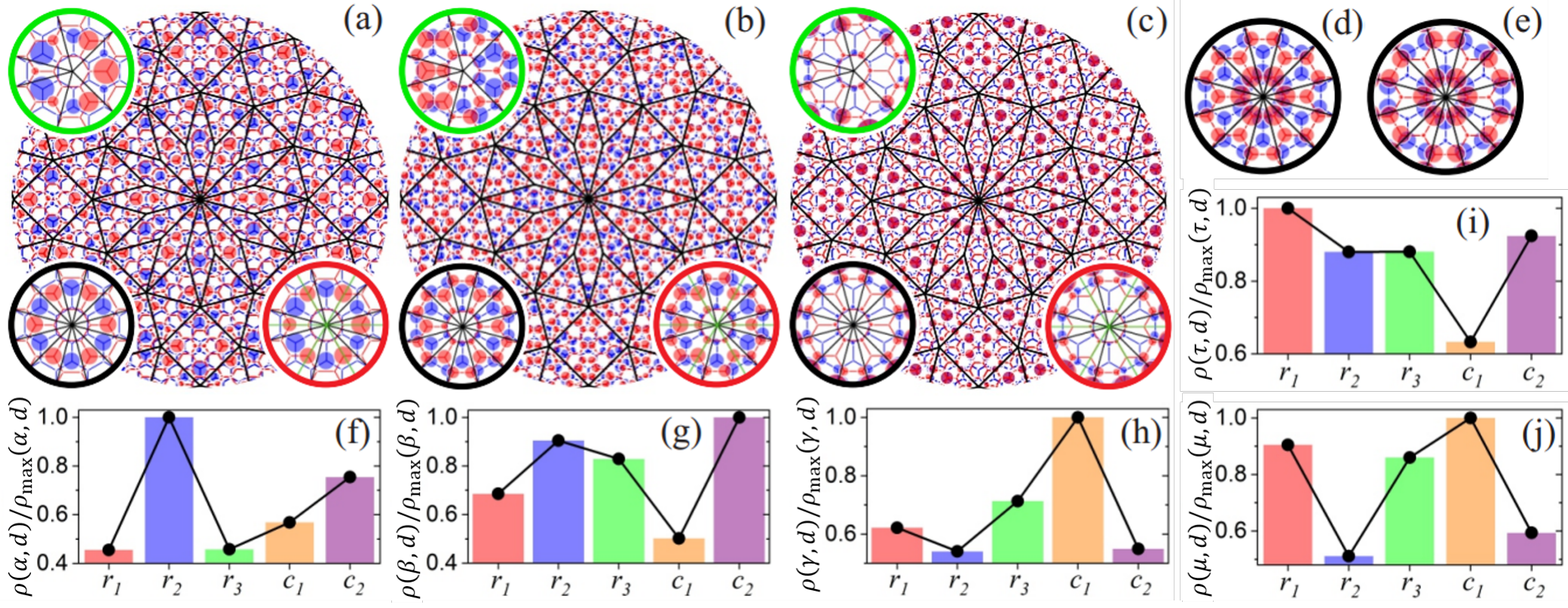}
\caption{
Real-space images of emergent states localization of a $30^{\circ}$-TBG system.
(a)-(e) correspond to the peaks $\alpha$, $\beta$, $\gamma$, $\tau$, $\mu$ marked in Figure \ref{fig:Fig_05}b, respectively.
Black lines represent the third inflation of Stampfli tiles used in the first construction method.
The areas of the red and blue disks are proportional to the LDOS at a particular energy in the top and bottom layers, respectively.
The black circles of (d), (e), and the insets of (a)-(c) are the zoom-in structures in relation to the central point that coincides with the twist axis, with perfect 12-fold symmetry.
The red and green circles of the insets of (a)-(c) are the zoom-in structure in relation to the central point formed by the vertices of different Stampfli tiles.
Bar graphs of (f)-(e) correspond to the DOS energy peaks $\epsilon_{i}$  with $i = \{ \alpha ,\beta ,\gamma ,\tau ,\mu \}$, respectively.
The histograms relate the LDOS at different radial distances $d=\{ r_{1},r_{2},r_{3},c_{1},c_{2} \}$, indicated in Figure \ref{fig:Fig_02}.
}
\label{fig:Fig_06}
\end{figure*}

Figures \ref{fig:Fig_06}(a)-(e) depict real-space images 
of emergent localized state of a $30^{\circ}$-TBG system, corresponding  to the peaks $\alpha$, $\beta$, $\gamma$, $\tau$, and $\mu$ of DOS singularities in the valence band, respectively.
The sizes of the red (blue) circles are proportional to the LDOS at the top (bottom) layer site at which the circle is centered.
The peak energies $\alpha$, $\beta$, $\gamma$, $\tau$ and $\mu$ are approximately $-1.163$~eV, $-0.949$~eV, $-0.702$~eV, $-0.842$~eV and $-0.44$~eV, respectively.
All emergent states exhibit a non-periodic dodecagonal pattern and can be spatially represented by a quasicrystal lattice model at the third inflation of the Spampfli tiles.

Figures \ref{fig:Fig_06}(f)-(j) present histograms illustrating the spatial distribution of states corresponding to the energy peaks $\epsilon_{i}$ with $i = \{ \alpha ,\beta ,\gamma ,\tau ,\mu \}$, respectively.
We analyze the variation of emergent localized states along the spacial radial distances, $d=\{ r_{1},r_{2},r_{3},c_{1},c_{2} \}$, as indicated in Figure \ref{fig:Fig_02}, where the vertices of the Stampfli tiles serve as central points.
Each histogram shows the average LDOS for the energy peak $\epsilon_{i}$, $\rho (\epsilon_{i},d)=\sum_{j\in d} \rho_{j}(\epsilon_{i})$, where $j$ corresponds to the atomic sites located at a distance $d$. The average LDOS is normalized by its maximum value, $\rho_{\rm max} (\epsilon_{i},d)$.

The black circles in the insets of Figs.~\ref{fig:Fig_06}(a)-(c) and in Figs.~\ref{fig:Fig_06}(d) and \ref{fig:Fig_06}(e) correspond to a zoom-in view of the structures of the central point regions formed by exclusively by Stampfli rhombuses associated with the energy peaks $\epsilon_{i}$ with $i=\{ \alpha,\beta,\gamma,\tau,\mu \}$, respectively. 
The distribution of the LDOS in each peak, $\rho_{j} (\epsilon_{i})$, exhibits a perfect 12-fold symmetry.
In the insets of Figs.~\ref{fig:Fig_06}(a)-(c), the red and green circles identify regions corresponding to dodecagonal centers used in the second construction method and the vertices of different Stampfli tiles used in the first construction method, respectively.
While the red circles do not locally preserve 12-fold symmetry, their distribution is close to the latter. 
However, the green circles deviate significantly from 12-fold symmetry,
displaying an imbalance due to the different symmetry groups of the Stampfli tiles.

We observe that the average LDOS at specific radial distances, $\rho (\epsilon_{i} ,d)$, in regions with strong imbalances resemble those in the regions with perfect 12-fold symmetry.

Figures \ref{fig:Fig_06}(a) and \ref{fig:Fig_06}(f) demonstrate that the $\alpha$-peak states are highly localized on the second ring of carbon atoms closest to the center points, at a radial distance $r_{2}$, with $\rho_{\rm max} (\alpha,r_{2})=0.203$~a.u.
Interestingly, these second-ring atomic sites can be considered as opposite of dimer sites, due to their significant spacing.
 
Figures \ref{fig:Fig_06}(b) and \ref{fig:Fig_06}(g) reveal a quasi-homogeneous  emergent states distribution for the $\beta$-peak,  with $\rho_{\rm max} (\beta,c_{2})=0.119 $~a.u., indicating less less localization at radial distance $c_{1}$.
Figures \ref{fig:Fig_06}(i) and \ref{fig:Fig_06}(j) show that the peaks $\tau$ and $\mu$ peaks also exhibit quasi-homogeneous emergent states distributions, with $\rho_{\rm max} (\tau,r_{1})=0.089$~a.u. and $\rho_{\rm max} (\mu,c_{1})=0.062$~a.u., respectively.
While the $\tau$-peak states are less localized at $c_{1}$, the $\mu$-peak states are less localized at $r_{2}$ and $c_{2}$.

Figure \ref{fig:Fig_06}(h)  illustrates that the $\gamma$-peak  states are highly localized at a radial distance $c_{1}$, coinciding with the dimer sites, with $\rho_{\rm max} (\gamma,c_{1})=0.102$~a.u.
In Figure \ref{fig:Fig_06}(c), the largest localizations are indicated by the purple disks, which correspond to overlapping red and blue disks representing the top and bottom layers, respectively.

\subsection{Suppression of the DOS}
\label{Suppression of the DOS}

We will now focus on the electronic density of states (DOS) suppression reported in Refs.~\cite{Yao2018, Yan2019}. 
Our goal is to use the geometric model introduced in Section~\ref{sec:atomic_structure} to understand the spatial distribution of the electronic states corresponding to the suppressed DOS.

\begin{figure}[h]
\centering
\includegraphics[width=0.7\columnwidth]{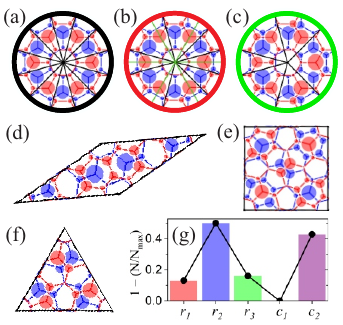}
\caption{
Emergent suppression strength in the LDOS of a $30^{\circ}$-TBG system.
The areas of the red and blue disks are proportional to the suppression strengths in the top and bottom layers, respectively.
(a)-(c) Zoom-in of the structures close to the twist axis, in the dodecagonal centers and in the vertices of different Stampfli tiles, respectively.
(d)-(f) represent the internal structures of the Stampfli tiles used in the first construction method.
The histogram in (g) relates the LDOS at different radial distances, indicated in Figure \ref{fig:Fig_02}.
}
\label{fig:Fig_07}
\end{figure}

Figures \ref{fig:Fig_07}(a)-(f) show the emergent suppressions in the LDOS of the $30^{\circ}$-TBG systems.
These suppressions are quantified by $S_{p}\equiv 1 - \rho (\epsilon_{i},d)/\rho_{\rm max} (\epsilon_{i},d)$ at the energy $\epsilon_{i}=-0.536$~eV, that lies between the $\gamma$ and $\mu$-peak energies.
A larger $S_{p}$ value indicates a greater suppression of the LDOS, suggesting stronger interlayer coupling between the graphene sheets.
The minimum and maximum LDOS values correspond to $\rho_{\rm min} (\epsilon_{i},r_{2})=0.019$~a.u. and $\rho_{\rm max} (\epsilon_{i},c_{1})=0.05$~a.u., respectively.

In Figure~\ref{fig:Fig_07}(a), the black circle corresponds to a zoomed-in view of the structure close to the twist axis.
The real-space distribution of suppression strengths exhibits a perfect 12-fold symmetry.
The red circle, Fig.~\ref{fig:Fig_07}(b), which corresponds to the dodecagonal centers, shows a LDOS distribution that only slightly deviates from 12-fold symmetry. 
In contrast, the green circle, corresponding to the vertices of different Stampfli tiles, exhibits a much stronger imbalance in the second ring closest to the center, violating the 12-fold symmetry.
Figs.~\ref{fig:Fig_07}(d)-(f) illustrate the distributions of suppression strengths within the internal structures of Stampfli tiles, exhibiting double, triple and quadruple symmetry. 
Fig.~\ref{fig:Fig_07}(g) reveals that the emergent suppression strengths are highly localized at radial distances $r_{2}$ and $c_{2}$.

\subsection{Landau levels spectrum}
\label{Landau levels spectrum}

Let us now perform a numerical investigation of the LL spectra in $30^{\circ}$-TBG systems and compare these results with experimental observations \citep{Yan2019}.
For that purpose, we compute the DOS, $\rho(\epsilon)$, by integrating the LDOS, $\rho_{j}(\epsilon)$, over the atomic sites within a Stampfli unitary structure.

\begin{figure}[h]
\centering
\includegraphics[width=0.95\columnwidth]{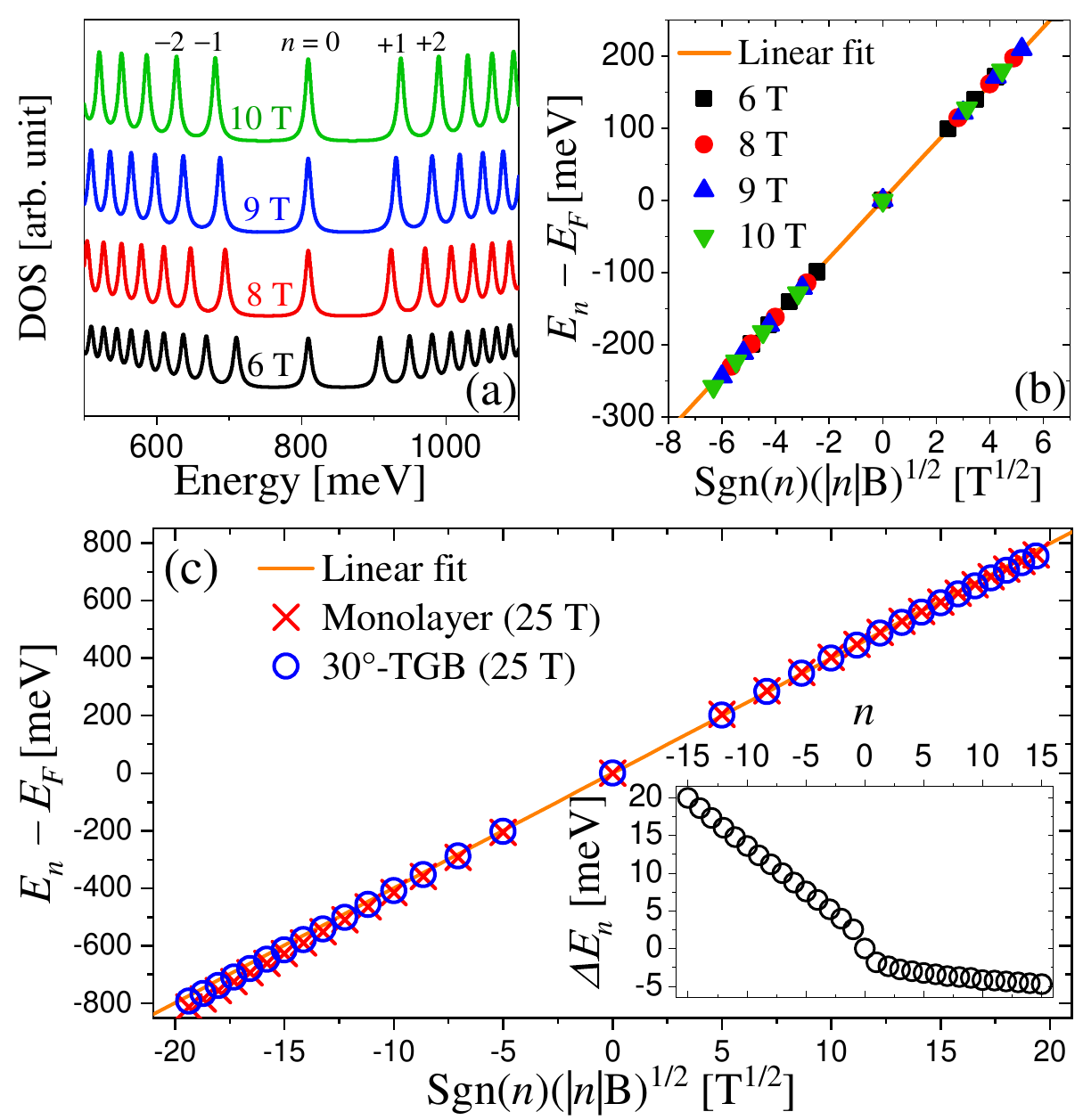}
\caption{
(a) DOS (in arbitrary units) as a function energy $E$ (in meV) of a pristine $30^{\circ}$-TBG system for magnetic fields of $B=6$, $8$, $9$ and $10$~T.
(b) LL peak energies (in meV) as a function of $\sqrt{\vert N\vert B}$, where $N$ is the LL index.
The symbols correspond to the peaks in (a), and the solid line shows the linear fit with the experimental data of Ref. \cite{Yan2019}.
(c) Comparison between the LL spectrum versus ${\rm Sgn}(N)(\vert N \vert B)^{1/2}$ for $30^{\circ}$-TBG and monolayer systems for $B=25$~T.
The inset shows the deviation $\Delta \epsilon_{N} = \epsilon^{\rm 30^{\circ}TBG}_{N} - \epsilon^{\rm G}_{N}$ (in meV) as a function the LL index $N$.
}
\label{fig:Fig_08}
\end{figure}

Figure \ref{fig:Fig_08}(a) shows a series of DOS for a pristine $30^{\circ}$-TBG system subjected to different perpendicular magnetic fields $B$.
The LLs are broadened using $\eta =5$~meV \citep{Vidarte2022}. 
The curves are offset in the $y$-axis for clarity.
The spectra consist of a sequence of well-defined LL peaks labeled by the index $N$ with increasing field.
Both the LL peak spacing and LL peak heights (related to the LL degeneracy) grow with increasing magnetic fields \citep{Goerbig2011}.

The LL energies, $\epsilon_{N}$, are determined by using a smaller regularization parameter, $\eta =0.1$~meV, and fitting the computed DOS of each LL peak with a Lorentzian distribution.
Figure \ref{fig:Fig_08}(b) shows that our results agree with the relation $\epsilon_{N}-\epsilon_{F}={\rm Sgn}(N)v_{F}\sqrt{2e\hslash\vert N\vert B}$, where $v_{F}$ the Fermi velocity, that is unique to massless Dirac fermions.
For clarity, we subtract the Fermi energy $\epsilon_{F}$, measured at the energy of the charge neutrality point.
Fitting both electron and hole branches of the experimental data yields a Fermi velocity of $v_{F}=(1.100 \pm 0.001)\times 10^{6}$~m/s, as reported in Ref. \cite{Yan2019}.
In summary, our numerical results show a good agreement with the STS measurements, confirming massless Dirac fermions in $30^{\circ}$-TBG systems.

Let us compare the $30^{\circ}$-TBG LL energies with those of graphene monolayers.
Figure \ref{fig:Fig_08}(c) shows $\epsilon_{N}$ for $30^{\circ}$-TBG and graphene monolayer systems as a function of $(\vert N\vert B)^{1/2}$ for $25$~T.
We verify that for low energies, corresponding to $\vert N\vert \lesssim 20$, the LL energies show, to very good accuracy, a linear dependence with $(\vert N\vert B)^{1/2}$.
Thus, we conclude that the electronic properties of the decagonal quasicrystal in $30^{\circ}$-TBG behave like a system with two separated monolayer graphene in the low-energy region.

Finally, let us examine the large-$\vert N\vert$ limit, where one expects the LLs energies to deviate from the linear behavior \citep{Goerbig2011,Vidarte2022}.
In the case of $30^{\circ}$-TBGs, these deviations arise due to long-range hopping terms that fit the nonlinear features of the dispersion relation \citep{Vidarte2022,Plochocka2008}.
The inset of the Figure \ref{fig:Fig_08}(c) gives the LL energy difference between $30^{\circ}$-TBG and graphene monolayer systems, $\Delta \epsilon_{N} = \epsilon^{\rm 30^{\circ}TBG}_{N} - \epsilon^{\rm G}_{N}$.
We observe a notable discrepancy in the valence band of $\sim 20$~meV for $N=-15$.
Nonetheless, the discrepancy is relatively small, reflecting the similarity of the DOS in the $30^{\circ}$-TBG and graphene systems at $B=0$.
At higher energies, near  $\epsilon\approx -1$~eV where the electronic states are localized (at $B=0$), we anticipate larger deviations.
However, given that the LLs spacing at such energies is $\sim 10$~meV at $B=25$~T,
experimental detection of these effects is challenging due to the broadening caused by the intrinsic disorder and the significant excitation and/or doping required.

\section{Conclusions and discussion}
\label{sec:concusion}

We have studied the local spectral properties and the emergence of real-space electronic localization in dodecagonal quasicrystal structures of $30^{\circ}$-TBG systems.
The HHK method's efficiency in calculating local spectral functions without relying on periodic boundary conditions outperforms atomistic calculations \cite{Moon2019} by orders of magnitude.
Furthermore, our findings exhibit excellent agreement with the localized states reported in Ref.~\cite{Moon2019} and the experimental density of states observed in Refs.~\cite{Yao2018, Yan2019}.

Our results indicate that the electronic structure of $30^{\circ}$-TBG system can be accurately calculated using only one of Stampfli tiles at the third inflation level, despite the observation that the spectral functions of the three Stampfli tiles are similar but not identical.
The spectral functions across the occupied electronic band reveal patterns characteristic of the dodecagonal quasicrystal, influenced by the varying number of carbon atom sites in the internal crystalline structure of the Stampfli tiles.

We developed a simple geometric model, applicable to the two lattice construction methods, to analyze the LDOS features in $30^{\circ}$-TBG systems. 
The emergent states associated with the singularities in the valence band are all identified with the Stampfli tiles, primarily those with imbalances in the non-perfect rings centered on the polygon vertices.
Perfect and near-perfect rings are identified with decagonal centers that exhibit emergent localization measures with pronounced 12-fold symmetry and optimal balance.

We find that the most pronounced effects of interlayer coupling between two graphene layers of the $30^{\circ}$-TBG system are localized in the second ring centered on the polygon vertices ($r_2$) and in the centers of the Stampfly squares ($c_2$).
This observation suggests a significant suppression of the LDOS at the intersection between the original and mirrored Dirac cones.

As an outlook, the efficiency of our method enables the analysis of twisted multilayer graphene systems \cite{Brzhezinskaya2021, Kononenko2022}, both commensurate and incommensurate, which involve a significantly larger number of atomic sites. 
Of particular interest are the recently observed moiré quasicrystals in twisted trilayer graphene  \cite{Hao2024}, which present a promising direction for further development of our work.

\acknowledgements

This work was partially supported by the Brazilian Institute of Science and Technology (INCT) in Carbon Nanomaterials and the Brazilian agencies CAPES, CNPq, FAPEMIG, and FAPERJ.


\bibliography{30TBG}


\end{document}